# Electrical Control of Large Rashba Effect in Oxide Heterostructures

Yan Song[1], Dong Zhang[2,3], Ben Xu[1,*], Kai Chang[2,3], Ce-Wen Nan[1,*]

[1]*School of Materials Science and Engineering, and State Key Laboratory of New Ceramics and Fine Processing, Tsinghua University, Beijing 100084, China*

[2]*SKLSM, Institute of Semiconductors, Chinese Academy of Sciences, P.O. Box 912, Beijing 100083, China*

[3]*Center for Excellent in Topological Quantum Computation, University of Chinese Academy of Sciences, Beijing 100190, China*

E-mail: xuben@mail.tsinghua.edu.cn, cwnan@mail.tsinghua.edu.cn

# Abstract

Large Rashba effect efficiently tuned by an external electric field is highly desired for spintronic devices. Using first-principles calculations, we demonstrate that large Rashba splitting is locked at conduction band minimum in ferroelectric Bi(Sc/Y/La/Al/Ga/In)$O_3$/PbTi$O_3$ heterostructures where the position of Fermi level is precisely controlled via its stoichiometry. Fully reversible Rashba spin texture and drastic change of Rashba splitting strength with ferroelectric polarization switching are realized in the symmetric and asymmetric heterostructures, respectively. By artificially tuning the local ferroelectric displacement and the orbital hybridization, the synergetic effect of local potential gradient and orbital overlap on the dramatic change of splitting strength is confirmed. These results improve the feasibility of utilizing Rashba spin-orbit coupling in spintronic devices.

Spin-orbit (SO) coupling, a relativistic effect that links spin and orbital angular momentum of electrons, offers opportunity for electrical control of spin degree of freedom [1, 2], which is appealing for spintronic applications due to the low consumption power [3]. In crystals without an inversion center, the gradient of crystal potential produces a momentum-dependent SO field called Rashba field [4]. This field couples with electron spin, splits the electronic energy bands and gives a unique spin-momentum locking which means charge current directly takes effect on spin. This property provides advantage in electrical manipulation of the magnetic state of matter [5, 6] and enables a variety of novel functionalities [7].

Rashba effect was at first investigated at the interfaces of semiconductor heterostructures [8] and metal surfaces [9, 10] but the splitting is rather small. Later on, so-called giant Rashba splitting was observed in Bi/Ag(111) surface alloy (with a splitting of 200 meV) [11] and bulk material BiTeI (a splitting of 100 meV) [12] using both angular-resolved photoemission spectroscopy and first-principles method. The Rashba splittings in these systems come from their structural asymmetry and are hardly tuned by external electric fields. In principle, the spin texture is controllable and switchable via an electric field if the materials possess sizable Rashba SO splitting and ferroelectricity simultaneously [13, 14]. The ferroelectric polarization which can be switched by an external electric field defines the sign of potential gradient, thus the Rashba spin splitting. Bulk GeTe is the prototype of ferroelectric Rashba semiconductor (FRS) and also the only FRS system investigated experimentally so far [15, 16]. Soon hexagonal KMgSb, LiZnSb, LiBeBi, NaZnSb, LiCaBi [17] and $BiAlO_3$ [18] are theoretically proposed as FRSs. $SiBiO_3$ is also a FRS candidate but a very large compressive strain is needed to stabilize the ferroelectric phase [19].

In fact, the FRS is also challenging for applications due to the dilemma between electric conduction (Rashba effect) and ferroelectric polarization being switched by gate voltage at the same time. For example, GeTe samples tend to form Ge vacancies and show considerable conductivity causing problem in switching ferroelectric state in conducting material [20, 21]. Then composite structures were designed to overcome this dilemma, where insulating ferroelectric switcher and conducting Rashba channel are in separated regions. A Bi adlayer on ferroelectric $BaTiO_3$ was first examined by first-principles calculations, however, the polarization of $BaTiO_3$ had negligible tunability on the Rashba splitting due to the weak interaction [22]. The Rashba effect was then found to be switchable in the proposed $BaTiO_3$/$Ba(Os,Ru,Ir)O_3$ heterostructures [23]. The switching of ferroelectric polarization is based on the network of oxygen octahedral. However, the Rashba bands held by 4*d* and 5*d* electrons in Os/Ru/Ir are 1 eV below the Fermi level hence hardly contribute to the conductivity. In the present article, practical switch of Rashba effect is demonstrated in Bi$M$O$_3$/PbTiO$_3$ ($M$=Sc, Y, La, Al, Ga, In) heterostructures. Bi *p* orbital with large Rashba splitting locates at the conduction band minimum (CBM) and the position of Fermi level can be precisely predicted.

First-principles calculations were carried out using VASP [24, 25] with the generalized gradient approximation by Perdew, Burke and Ernzerhof [26]. Kohn-Sham single particle wavefunctions are expanded in plane wave basis set with a kinetic energy cutoff at 600 eV. A 10×10×2 *k*-point grid centered at the Γ point is used. Lattice constants are fixed at the experimental value of PbTiO$_3$ [27]. All the atomic positions are free to relax during the optimization until the energy difference and force converge to $10^{-5}$ eV and 0.01 eV/Å, respectively. Coulomb interaction $U$=3 eV is added on Ti 3*d* orbital. Spin texture is plotted with PyProcar code [28]. Density of states is plotted using Sumo [29].

The designed structure is based on ferroelectric $AB$O$_3$ perovskite, where we change the *A*-site cation in the middle layers to accommodate large Rashba splitting and robust ferroelectricity. Meanwhile, the Rashba bands are kept at CBM via selecting the suitable *B*-site elements. PbTiO$_3$ is a common ferroelectric with tetragonal *P4mm* symmetry below 766 K [30]. Alternating [PbO] and [TiO$_2$] layers are stacked along [001] direction. The ferroelectric structure is characterized by the relative displacements of the cations and anions. Switchable dipole moment is valid and has been used in many important fields [31-33]. Pb atom was substituted by Bi in the middle layers aiming to utilize the Bi *p* orbital to hold large Rashba effect [11, 22, 34]. For *B*-site cation, Ti and other metal elements with partially filled *d* shell are detrimental because *d* state will be in the vicinity of Fermi level leaving Bi *p* orbital out of the picture. Also it's hard to anticipate the position of Fermi level with high- and low-spin-state occupation. To eliminate the interference of *d* electrons, Sc ($3d^14s^2$), Y ($4d^15s^2$), La ($5d^16s^2$), Al ($3s^2p^1$), Ga ($4s^2p^1$) and In ($5s^2p^1$) are ideal for *B*-site cation. They present empty or full *d* shell in Bi$M$O$_3$, so the Bi *p* orbital is expected as the lowest conduction band.

Although cubic $Pm\bar{3}m$ or tetragonal *P4mm* is not the lowest energy structure for bulk Bi(Sc/Y/La/Al/Ga/In)O$_3$ [35], they are expected to grow on PbTiO$_3$ and stabilize a tetragonal ferroelectric phase. According to computational works [36], Bi(Sc/Y/Al/Ga/In)O$_3$ with cubic $Pm\bar{3}m$ structure have considerable negative value of formation energy (-1.3~-2.5 eV) and they tend to form ferroelectric tetragonal phase due to the moderate epitaxial strain from PbTiO$_3$. BiLaO$_3$ has a rhombohedral $R3c$ structure with formation energy of -2.7 eV [36].

Moreover, [PbO] and [TiO$_2$] layers are charge neutral and tend to form doping-type interfaces when contacting with [BiO]$^+$ and [$M$O$_2$]$^-$. For example, with three layers of [BiO]$^+$ and two layers of [$M$O$_2$]$^-$, one electron is expected to be confined in the Bi$M$O$_3$ region so the Fermi level would cross the bottom conduction band. Similarly, Fermi level would locate in the semiconducting gap or cross the top valence band when the layer number of [BiO]$^+$ is equal to or one less than that of [$M$O$_2$]$^-$, respectively. The first two cases are of most concern. Since substituting $M$ by Sc, Y, La, Al, Ga and In yields a similar effect, we just use BiInO$_3$/PbTiO$_3$ as an example for illustration since bulk BiAlO$_3$ has been predicted to present ferroelectric tetragonal phase [18] and BiInO$_3$ has similar electronic configuration but with larger SO strength. Band structures for the other elements can be found in Fig. S1 in Supporting Information.

Fig. 1(a) and (b) show two BiInO$_3$ sandwiched by four PbTiO$_3$ layers with an extra [BiO]$^+$ in the middle with two different ferroelectric polarizations. These structures have inversion centers in the paraelectric phase, so the potential gradient is only introduced by the ferroelectric displacements and were labeled as symmetric models. As expected, the Fermi level crosses the bottom conduction band which is mainly contributed from Bi $p$ orbital. The magnified band structures below Fermi level around Γ point are given in Fig. 1(c) and (d). The second lowest pair of bands shows Rashba splitting of 7 meV along Γ–M and 6 meV along Γ–X as listed in Table I, compared with other ferroelectric Rashba systems proposed in the previous theoretical studies. The splitting is in the same order of magnitude as the splitting in bulk BiAlO$_3$ [18], SrBiO$_3$ [19], KMgSb [17] and BaOsO$_3$/BaTiO$_3$ heterostructure [23]. The advantage of the proposed structure here is that the Rashba bands are in the vicinity of Fermi level and the insulating PbTiO$_3$ acting as ferroelectric switcher can be excluded from the conducting channel which is practical for gate voltage manipulation. The potential gradient and the spin texture change sign when the polarization reverses, while the band structure, i.e., the Rashba splitting strength, remains the same. This can be found from the three projected components of spin in Fig. 1(e) and (f), where purely two-dimensional Rashba texture presents. Spins are along tangential directions in the $k_x$-$k_y$ plane and the out-of-plane component is zero.

As mentioned, when the layer number of [BiO]$^+$ is equal to that of [InO$_2$]$^-$, the system remains semiconducting. Fig. 2(a) and (b) show the asymmetric models with one layer of BiInO$_3$ sandwiched by four layers of PbTiO$_3$. The lowest pair of conduction band is mainly from Bi $p$ orbital. The splitting is 13 meV along Γ–M and 14 meV along Γ–X in [001] polarization [Fig. 2(a)]. With reversed polarization, splitting at Γ point almost quenches and reduces to 0.2 meV [Fig. 2(b)]. In isotropic two-dimensional electron gas model, Rashba parameter $\alpha_R = 2E_R/k_R$ is used to quantify the Rashba SO strength where $E_R$ is the Rashba splitting energy and $k_R$ is the momentum offset. The $\alpha_R$ experiences drastic change from 0.929 (Table I) to 0.014 eV·Å when polarization is reversed. Fig. 2(c) and (d) are the band contour images as functions of $k_x$ and $k_y$ at certain energy relative to the crossing point of Rashba bands. The out-of-plane components are zero hence not shown in Fig. 2(c) and (d). Rashba characteristic of the inner and outer contours are apparent. Their shapes are nearly circular. The anticlockwise direction of spin in the inner circle [Fig. 2(c)] changes to clockwise [Fig. 2(d)] by reversing the polarization. Increasing energy from the crossing point, circles show square intensity, reflecting the four-fold in-plane rotational symmetry.

The plausible explanation for the change of $\alpha_R$ with reversing the polarization is the intrinsic structural asymmetry as the two adjacent layers to [BiO]$^+$ is [InO$_2$]$^-$ and [TiO$_2$]. The [InO$_2$]$^-$ presents nominate negative charge of $1e$ while the [TiO$_2$] is neutral. So it favors the dipole moment in [BiO]$^+$ layer along [001] than [00$\bar{1}$] direction which means the ferroelectric displacement in [BiO]$^+$ layer, i.e., the local potential gradient, would be different when polarization changes. Fig. 3(a) is the partial charge density for the lowest pair of conduction band around Γ point. The bands are mainly from Bi and Ti orbital. The purple shadow indicates the magnitude of displacement in [BiO]$^+$ layer. The displacement is 0.93 Å in [001] much larger than 0.46 Å in [00$\bar{1}$]. The upper and lower black lines filled with gray in Fig. 3(b) give the magnified Rashba bands of [001] and [00$\bar{1}$] model, respectively, but enlarging the Bi-O displacement in [00$\bar{1}$] model back to 0.93 Å could not increase the Rashba splitting. Hence, the local potential gradient in [BiO]$^+$ layer is not the only factor causing the drastic change of $\alpha_R$.

Besides, the polarization changes the orbital overlap significantly. As in Fig. 3(a), Bi is closer to O atoms in [InO$_2$]$^-$ layer in one polarization direction and [TiO$_2$] in the other. The hybridization between two orbitals is strongly dependent on their energy difference and localization extent. In density functional calculations, $3d$ orbital is much more sensitive to the on-site Coulomb interaction than $p$ orbital. The LDA+$U$ method implemented in VASP then becomes a practical method to artificially tune the localization of the adjacent Ti $3d$ orbital hence the hybridization between Ti and Bi states.

Fig. 3(c) and the upper two panels in (d) show the atom- and orbital-resolved density of states for [001] and [00$\bar{1}$] model, respectively. For [001], the CBM is only dominated by Bi $p_z$ orbital while for [00$\bar{1}$] Bi $p$ and Ti $d_{xy}$ states show major overlap. To find out the role of orbital overlap on Rashba SO coupling, we increase the energy of Ti $d$ states by artificially change the Coulomb $U$ of the nearest Ti 3$d$ orbital to 6 and 12 eV in [00$\bar{1}$] model. The energy of Ti 3$d$ orbital increases significantly while the Bi $p$ orbital is hardly influenced when the localization is strengthened, as shown in the lower panels in Fig. 3(d). Hence, the hybridization between these two orbitals can be effectively tuned. Red and blue dashed lines in Fig. 3(b) are the resulting Rashba bands. The splitting shows trivial relevance with the variation of $U$. Therefore, the contribution of structural asymmetry was considered by comparing the band structure for [00$\bar{1}$] model with both the artificially increased Bi-O displacement (0.93 Å) and $U$. As seen from the pink and green dashed lines, the bands show great tendency of reversing back to the splitting in [001] case. Hence, it is concluded that the synergetic effect of the ferroelectric displacement and the orbital hybridization modified by polarization are responsible for the drastic change of Rashba splitting. The latter effect was also observed in polar two-dimensional transition-metal dichalcogenides where the overlap between W $d_{z^2}$ and Se $p_z$ orbital dominates the Rashba SO strength [37].

The layer-dependent electronic properties (Fig. S2 and S3) and the band structures using more accurate hybrid functional HSE06 (Fig. S4) have also been checked. For both symmetric and asymmetric models, the number of Bi $p$ Rashba bands is equal to the layer number of [BiO]$^+$ but the relevant physics discussed above is not influenced. Since the PbTiO$_3$ bands locate further from the Fermi level and increasing its layer number neither changes the band gap nor affects the Rashba effect so the insulating ferroelectric can be thick enough to maintain polarization. Additionally the indirect band gap of the asymmetric model is increased from 0.74 to 1.68 eV by HSE06 but the characteristic of Rashba bands was marginally affected proving the GGA results above are valid.

In summary, we have performed relativistic first-principles calculations to investigate the Rashba effect in ferroelectric PbTiO$_3$/Bi(Sc/Y/La/Al/Ga/In)O$_3$ heterostructures. Our results show that the position of Fermi level can be precisely controlled by the stoichiometry of [BiO]$^+$ and [$M$O$_2$]$^-$. Fully reversible Rashba spin texture with ferroelectric polarization switching can be realized in symmetric models because the ferroelectric displacement alone determines the potential gradient. Reversible Rashba effect and dramatic change in splitting strength could be achieved in asymmetric models, which can be attributed to the intrinsic structural asymmetry. This electrically tuning of Rashba splitting on both strength and sign is appealing for spintronic and magnetoelectric devices.

## Acknowledgements

This work was supported by the National Natural Science Foundation of China (51790494 and 61788104).

# Table notes

**Table I.** Rashba energy $E_R$, momentum offset $k_R$ and Rashba parameter $\alpha_R$ of the symmetric and asymmetric PbTiO$_3$/BiInO$_3$ heterostructures and other ferroelectric Rashba systems in previous theoretical studies.

| System | $E_R$ (meV) | $k_R$ (Å$^{-1}$) | $\alpha_R$ (eV·Å) | Reference |
|---|---|---|---|---|
| PbTiO$_3$/BiInO$_3$ heterostructure (symmetric)[a] | 7 (6) | 0.028 (0.040) | 0.500 (0.300) | This work |
| PbTiO$_3$/BiInO$_3$ heterostructure (unsymmetric)[b] | 13 (14) | 0.028 (0.040) | 0.929 (0.700) | This work |
| Bulk BiAlO$_3$ (*R3c*) | 7 | 0.040 | 0.390 | [18] |
| Bulk BiAlO$_3$ (*P4mm*) | 9 | 0.030 | 0.740 | [18] |
| Bulk SrBiO$_3$[c] | <5 | — | 0.944 | [19] |
| Bulk KMgSb | 10 | 0.024 | 0.830 | [17] |
| Bulk LiZnSb | 21 | 0.023 | 1.820 | [17] |
| Bulk LiBeBi | 24 | 0.026 | 1.840 | [17] |
| Bulk NaZnSb | 31 | 0.024 | 2.580 | [17] |
| Bulk LiCaBi | 32 | 0.035 | 1.820 | [17] |
| Bi(111)/BaTiO$_3$ heterostructure[d] | 160 | 0.220 | 1.450 | [22] |
| BaOsO$_3$/BaTiO$_3$ heterostructure[e] | 4 | 0.043 | 0.186 | [23] |
| BaIrO$_3$/BaTiO$_3$ heterostructure[e] | 53 | 0.145 | 0.731 | [23] |
| BaRuO$_3$/BaTiO$_3$ heterostructure[e] | 16 | 0.128 | 0.250 | [23] |

[a] Splitting of the second lowest pair of conduction bands along Γ-M (Γ-X)

[b] Splitting of the lowest pair of conduction bands along Γ-M (Γ-X)

[c] Large compressive strain to stabilize ferroelectric phase

[d] Hardly change with substrate ferroelectric polarization

[e] Rashba bands far from Fermi level

**Figure captions**

**Fig. 1.** Band structure of symmetric PbTiO$_3$/BiInO$_3$ heterostructure with ferroelectric polarization along (a) [001] and (b) [00$\bar{1}$]. The blue circles represent Bi $p$ orbital. Rashba bands around Γ point below Fermi level are magnified in (c) and (d). Spin textures at Fermi level in $k_x$-$k_y$ plane are shown in (e) and (f). Spin projection is depicted as the color scale.

**Fig. 2.** Band structure of asymmetric PbTiO$_3$/BiInO$_3$ heterostructure with ferroelectric polarization along (a) [001] and (b) [00$\bar{1}$]. The blue circles represent Bi $p$ orbital. The lowest pair of conduction bands around Γ point is magnified. Spin textures at energy surfaces 0.01, 0.02, 0.03 and 0.2 eV above the crossing point are shown in (c) and (d). Spin projection is depicted as the color scale.

**Fig. 3.** (a) Partial charge density for the lowest pair of conduction bands of the asymmetric PbTiO$_3$/BiInO$_3$ around Γ point. The purple shadow indicates the magnitude of ferroelectric displacement. (b) Rashba splitting bands with different Coulomb interaction parameters. Upper and lower gray bands are initial cases with polarization along [001] and [00$\bar{1}$] direction, respectively. Atom- and orbital-resolved density of states with polarization along (c) [001] and (d) [00$\bar{1}$].

**Fig. 1**

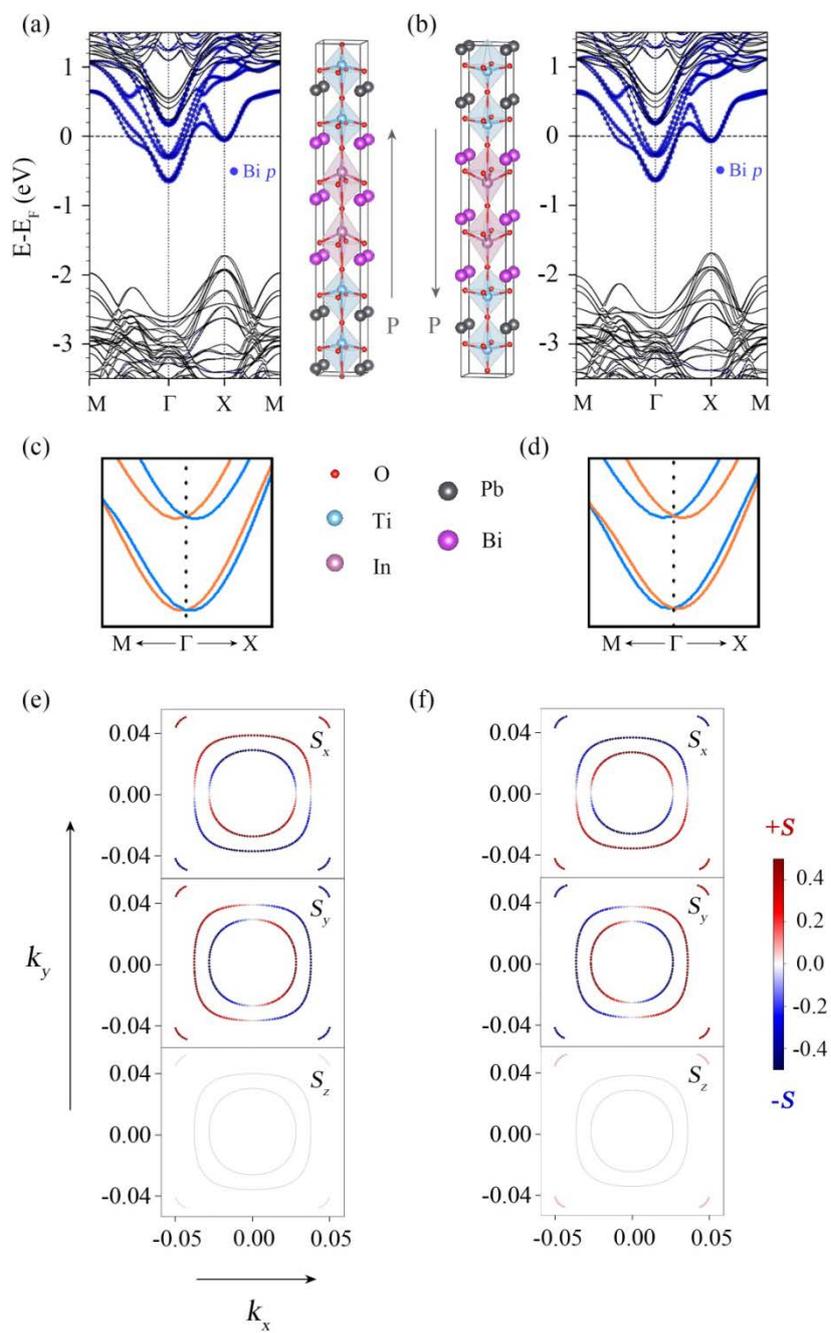

**Fig. 2**

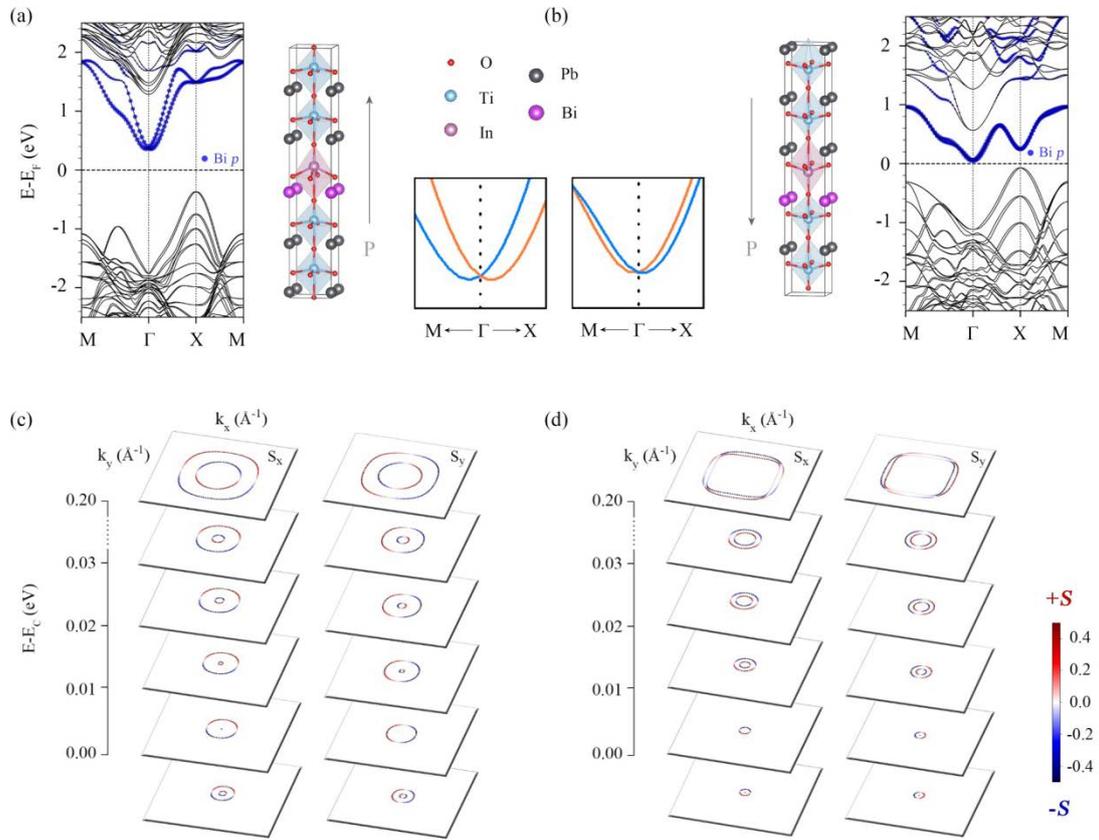

**Fig. 3**

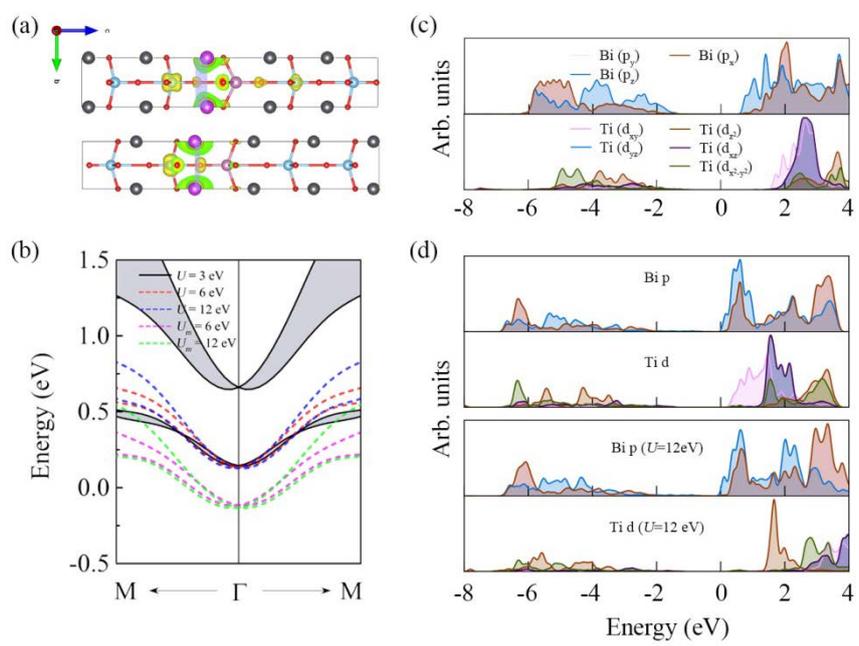

# References

1   Kuschel, T. and G. Reiss, Charges ride the spin wave, Nat. Nanotechnol., 2015. 10(1): 22-24.

2   Vaz, D.C., A. Barthélémy, and M. Bibes, Oxide spin-orbitronics: New routes towards low-power electrical control of magnetization in oxide heterostructures, Jpn. J. Appl. Phys., 2018. 57(9): 0902A4.

3   Garcia, V., et al., Ferroelectric control of spin polarization, Science, 2010. 327(5969): 1106-1110.

4   Rashba, E.I., Properties of semiconductors with an extremum loop. I. Cyclotron and "combinational" resonance in a magnetic field perpendicular to the plane of the loop, Fizika Tverdogo Tela, 1960. 2(6): 1224-1238.

5   Edelstein, V.M., Spin polarization of conduction electrons induced by electric-current in 2-dimensional asymmetric electron systems, Solid State Commun., 1990. 73(3): 233-235.

6   Datta, S. and B. Das, Electronic analog of the electro-optic modulator, Appl. Phys. Lett., 1990. 56(7): 665-667.

7   Manchon, A., et al., New perspectives for Rashba spin-orbit coupling, Nat. Mater., 2015. 14(9): 871-882.

8   Nitta, J., et al., Gate control of spin-orbit interaction in an inverted $In_{0.53}Ga_{0.47}As/In_{0.52}Al_{0.48}As$ heterostructure, Phys. Rev. Lett., 1997. 78(7): 1335-1338.

9   LaShell, S., B.A. McDougall, and E. Jensen, Spin splitting of an Au(111) surface state band observed with angle resolved photoelectron spectroscopy, Phys. Rev. Lett., 1996. 77(16): 3419-3422.

10  Popovic, D., et al., High-resolution photoemission on Ag/Au(111): Spin-orbit splitting and electronic localization of the surface state, Phys. Rev. B, 2005. 72(4): 045419.

11  Ast, C.R., et al., Giant spin splitting through surface alloying, Phys. Rev. Lett., 2007. 98(18): 186807.

12  Ishizaka, K., et al., Giant Rashba-type spin splitting in bulk BiTeI, Nat. Mater., 2011. 10: 521-526.

13  Di Sante, D., et al., Electric control of the giant Rashba effect in bulk GeTe, Adv. Mater., 2013. 25(4): 509-513.

14　Picozzi, S., Ferroelectric Rashba semiconductors as a novel class of multifunctional materials, Frontiers in Physics, 2014. 2(10): 1-5.

15　Liebmann, M., et al., Giant Rashba-type spin splitting in ferroelectric GeTe(111), Adv. Mater., 2016. 28(3): 560-565.

16　Rinaldi, C., et al., Ferroelectric control of the spin texture in GeTe, Nano Lett., 2018. 18(5): 2751-2758.

17　Narayan, A., Class of Rashba ferroelectrics in hexagonal semiconductors, Phys. Rev. B, 2015. 92(22): 220101.

18　da Silveira, L.G.D., P. Barone, and S. Picozzi, Rashba-Dresselhaus spin-splitting in the bulk ferroelectric oxide $BiAlO_3$, Phys. Rev. B, 2016. 93(24): 245159.

19　Varignon, J., J. Santamaria, and M. Bibes, Electrically switchable and tunable Rashba-type spin splitting in covalent perovskite oxides, Phys. Rev. Lett., 2019. 122(11): 116401.

20　Kolobov, A.V., et al., Local structure of crystallized GeTe films, Appl. Phys. Lett., 2003. 82(3): 382-384.

21　Giussani, A., et al., On the epitaxy of germanium telluride thin films on silicon substrates, Phys. Status Solidi B, 2012. 249(10): 1939-1944.

22　Mirhosseini, H., et al., Toward a ferroelectric control of Rashba spin-orbit coupling: Bi on $BaTiO_3$(001) from first principles, Phys. Rev. B, 2010. 81(7): 073406.

23　Zhong, Z.C., et al., Giant switchable Rashba effect in oxide heterostructures, Adv. Mater. Interfaces, 2015. 2(5): 1400445.

24　Kresse, G. and J. Furthmuller, Efficient iterative schemes for ab initio total-energy calculations using a plane-wave basis set, Phys. Rev. B, 1996. 54(16): 11169-11186.

25　Kresse, G. and D. Joubert, From ultrasoft pseudopotentials to the projector augmented-wave method, Phys. Rev. B, 1999. 59(3): 1758-1775.

26　Perdew, J.P., K. Burke, and M. Ernzerhof, Generalized gradient approximation made simple, Phys. Rev. Lett., 1996. 77(18): 3865-3868.

27　Glazer, A.M. and S.A. Mabud, Powder profile refinement of lead zirconate titanate at several temperatures. II. Pure $PbTiO_3$, Acta Crystallogr. B, 1978. 34(4): 1065-1070.

28　Herath, U., et al., PyProcar: A Python library for electronic structure pre/post-processing, arXiv:1906.11387, 2019.

29   Ganose, A.M., A.J. Jackson, and D.O. Scanlon, Sumo: Command-line tools for plotting and analysis of periodic ab initio calculations, Journal of Open Source Software, 2018. 3(28): 717.

30   Cohen, R.E., Origin of ferroelectricity in perovskite oxides, Nature, 1992. 358(6382): 136-138.

31   Stoica, V.A., et al., Optical creation of a supercrystal with three-dimensional nanoscale periodicity, Nat. Mater., 2019. 18(4): 377-383.

32   Dawber, M., et al., Tailoring the properties of artificially layered ferroelectric superlattices, Adv. Mater., 2007. 19(23): 4153-4159.

33   Sinsheimer, J., et al., Engineering polarization rotation in a ferroelectric superlattice, Phys. Rev. Lett., 2012. 109(16): 167601.

34   Koroteev, Y.M., et al., Strong spin-orbit splitting on Bi surfaces, Phys. Rev. Lett., 2004. 93(4): 046403.

35   Wang, H., et al., First-principles study of the cubic perovskites Bi*M*O$_3$ (*M*=Al, Ga, In, and Sc), Phys. Rev. B, 2007. 75(24): 245209.

36   Jain, A., et al., Commentary: The Materials Project: A materials genome approach to accelerating materials innovation, APL Materials, 2013. 1(1): 011002.

37   Yao, Q.-F., et al., Manipulation of the large Rashba spin splitting in polar two-dimensional transition-metal dichalcogenides, Phys. Rev. B, 2017. 95(16): 165401.

# Electrical Control of Large Rashba Effect in Oxide Heterostructures

Yan Song[1], Dong Zhang[2,3], Ben Xu[1,*], Kai Chang[2,3], Ce-Wen Nan[1,*]

[1]School of Materials Science and Engineering, and State Key Laboratory of New Ceramics and Fine Processing, Tsinghua University, Beijing 100084, China

[2]SKLSM, Institute of Semiconductors, Chinese Academy of Sciences, P.O. Box 912, Beijing 100083, China

[3]Center for Excellent in Topological Quantum Computation, University of Chinese Academy of Sciences, Beijing 100190, China

## Contents:

S1. Band structures of PbTiO$_3$/Bi(Al/Ga/In/Sc/Y/La)O$_3$ heterostructures

S2. Layer-dependent band structures of the symmetric models

S3. Layer-dependent band structures of the unsymmetric models

S4. Hybrid functional band structure of PbTiO$_3$/BiInO$_3$ heterostructure

## S1. Band structures of PbTiO$_3$/Bi(Al/Ga/In/Sc/Y/La)O$_3$ heterostructures

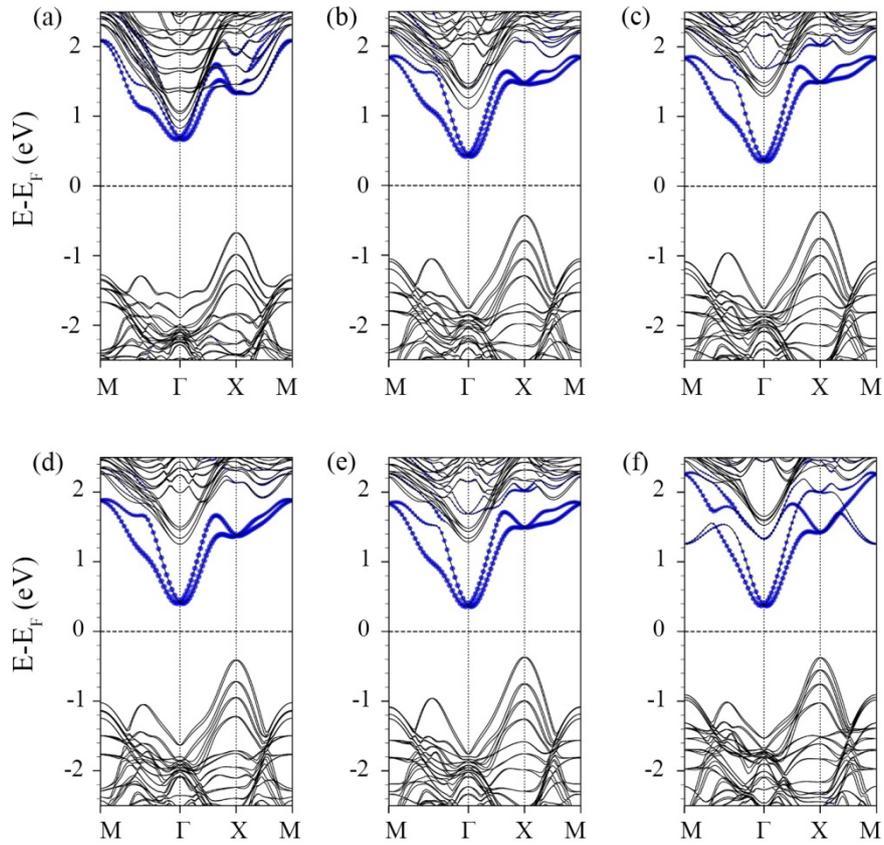

**Fig. S1.** Band structures of the unsymmetric PbTiO$_3$/Bi$M$O$_3$ with $M$ equals to (a) Al, (b) Ga, (c) In, (d) Sc, (e) Y and (f) La. The layer number of PbTiO$_3$ and Bi$M$O$_3$ are 4 and 1, respectively.

## S2. Layer-dependent band structures of the symmetric models

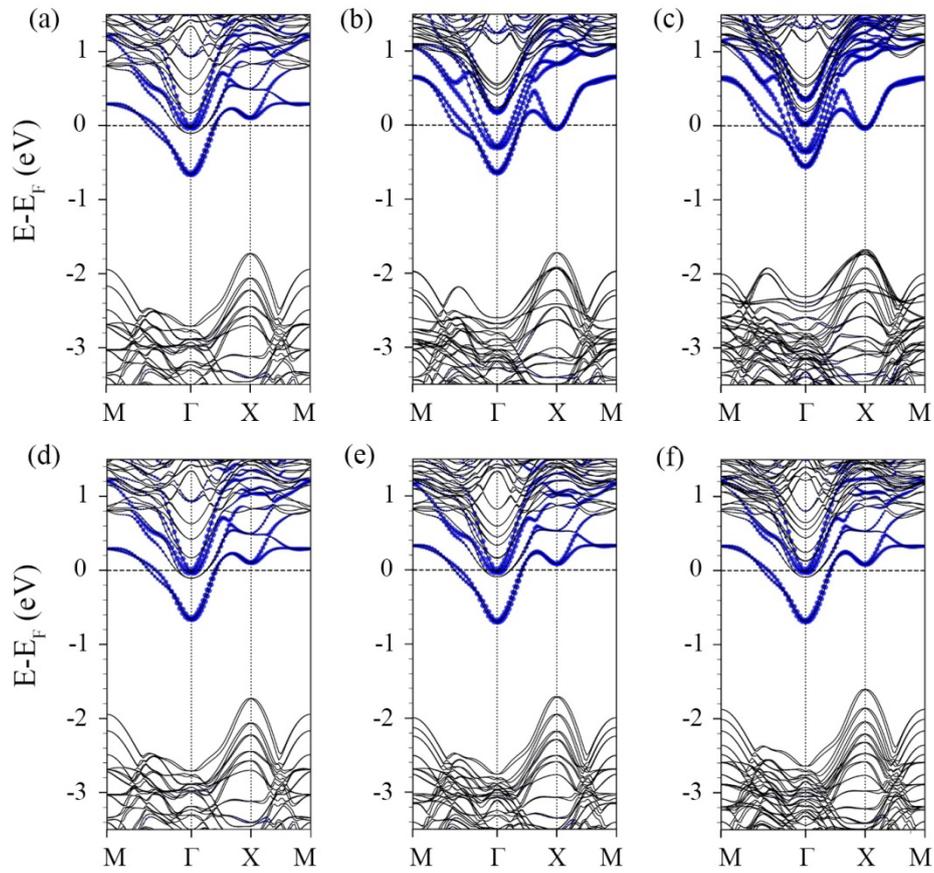

**Fig. S2.** Band structures of the symmetric PbTiO$_3$/BiInO$_3$ with (a) 1, (b) 2 and (c) 3 layers of BiInO$_3$ and 4 layers of PbTiO$_3$. Bands with 1 layer of BiInO$_3$ and (d) 4, (e) 5 and (f) 6 layers of PbTiO$_3$. The blue circles represent Bi $p$ orbital component.

## S3. Layer-dependent band structures of the unsymmetric models

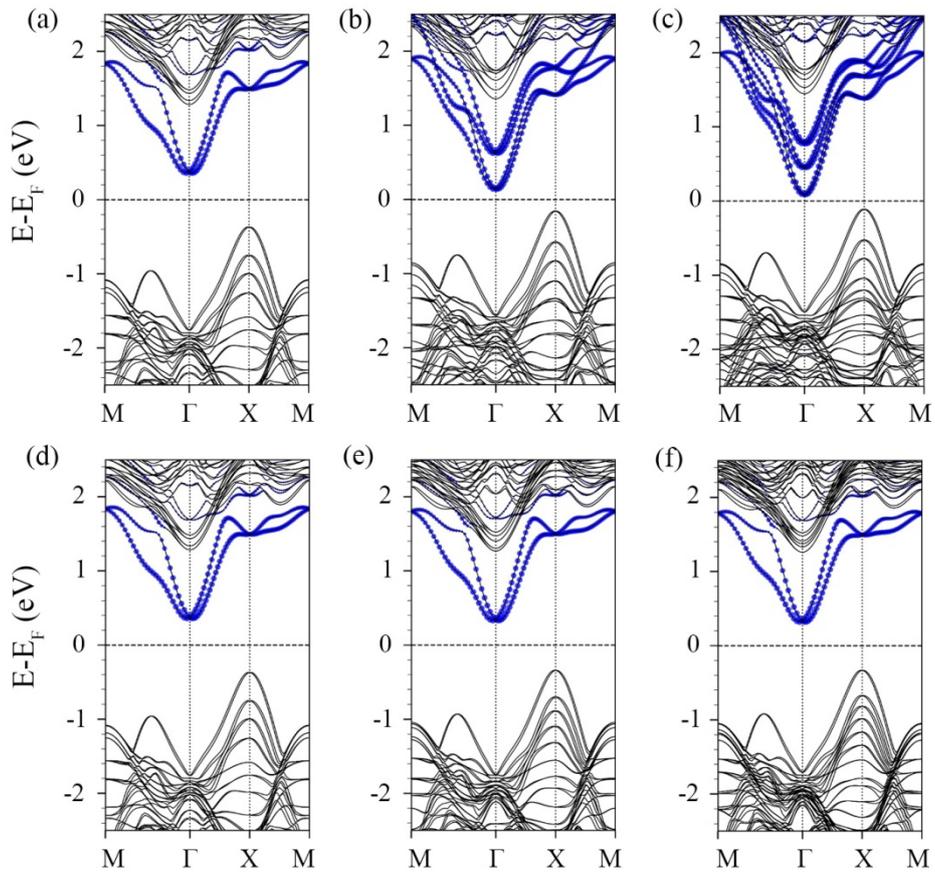

**Fig. S3.** Band structures of the unsymmetric PbTiO$_3$/BiInO$_3$ with (a) 1, (b) 2 and (c) 3 layers of BiInO$_3$ and 4 layers of PbTiO$_3$. Bands with 1 layer of BiInO$_3$ and (d) 4, (e) 5 and (f) 6 layers of PbTiO$_3$. The blue circles represent Bi *p* orbital component.

## S4. Hybrid functional band structure of PbTiO$_3$/BiInO$_3$ heterostructure

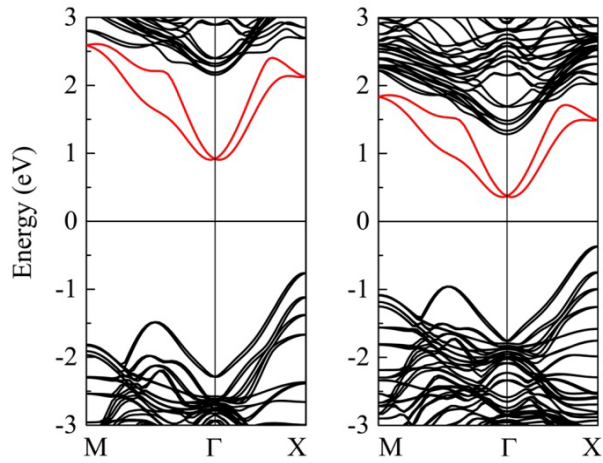

**Fig. S4.** Band structures of the unsymmetric PbTiO$_3$/BiInO$_3$ with (a) HSE06 and (b) PBE functional. The indirect band gap is increased from 0.74 to 1.68 eV by the hybrid functional while the Rashba dispersion relation is not affected.